\documentclass[conference,a4paper]{APSIPA2021}
\usepackage{multirow}
\usepackage{graphicx}
\usepackage{amssymb}
\usepackage{amsxtra}
\usepackage{threeparttable}

\usepackage{geometry}
\usepackage{caption}
\usepackage{subcaption}
\usepackage{url}
\usepackage{array}
\usepackage{booktabs}
\usepackage{balance}

\begin{document}

\title{Adversarial Speaker-Consistency Learning Using Untranscribed Speech Data for Zero-Shot Multi-Speaker Text-to-Speech}

\author{
\authorblockN{
Byoung Jin Choi,
Myeonghun Jeong,
Minchan Kim,
Sung Hwan Mun,
Nam Soo Kim
}

\authorblockA{
Department of Electrical and Computer Engineering and INMC, Seoul National University, Seoul, Korea \\
E-mail: \{bjchoi, mhjeong, mckim, shmun\}@hi.snu.ac.kr, nkim@snu.ac.kr Tel/Fax: +82-02-884-1824}

}

\maketitle

\begin{abstract}
Several recently proposed text-to-speech (TTS) models achieved to generate the speech samples with the human-level quality in the single-speaker and multi-speaker TTS scenarios with a set of pre-defined speakers. However, synthesizing a new speaker's voice with a single reference audio, commonly known as zero-shot multi-speaker text-to-speech (ZSM-TTS), is still a very challenging task. The main challenge of ZSM-TTS is the speaker domain shift problem upon the speech generation of a new speaker. To mitigate this problem, we propose adversarial speaker-consistency learning (ASCL). The proposed method first generates an additional speech of a query speaker using the external untranscribed datasets at each training iteration. Then, the model learns to consistently generate the speech sample of the same speaker as the corresponding speaker embedding vector by employing an adversarial learning scheme. The experimental results show that the proposed method is effective compared to the baseline in terms of the quality and speaker similarity in ZSM-TTS.
\end{abstract}

\section{Introduction}
The performance of the neural text-to-speech (TTS) models has dramatically improved in terms of the quality of the speech samples in a recent few years. While the most innovations in TTS field emerged around enhancing the quality of the single-speaker and the multi-speaker models which are trained with a pre-defined speaker set, the deployment in the field application requires TTS systems with various capabilities. One popular demand is an instant speaker adaptation to build a personalized TTS system. Personalized TTS aims to analyze and control the underlying speech factors to imitate the user's voice characteristics. Nonetheless, these factors are not rigorously defined in a scientific manner and generally known to be entangled which makes it difficult to control each component. In personalized TTS, the main objective is to adapt to a new speaker's voice characteristics with limited data available. To realize such demand, zero-shot multi-speaker TTS (ZSM-TTS), a sub-branched research under the umbrella of speaker adaptation, has gained an enormous attention from researchers recently. ZSM-TTS seeks to train a multi-speaker TTS model which can generate a speech sample of a new speaker's voice identity which was not present in the training dataset given a reference utterance without further finetuning the model.

Some previous works have tackled ZSM-TTS using a pre-trained speaker encoder for speaker verification to the existing TTS models \cite{Transfer-Tacotron}, \cite{zero-shot-tts}, \cite{SC-GlowTTS}, \cite{YourTTS}. Meanwhile, an effective style modeling method was proposed by \cite{GST}, where a bank of style vectors and their weights are learned in an unsupervised manner. On the other hand, \cite{Meta-StyleSpeech} exploits a meta-learning approach by utilizing an episodic training scheme with the phoneme and style discriminators. However, although these approaches focus on improving the speaker embedding extraction, conditioning scheme, and training method, they disregard the fact that the number of speakers in the current TTS training dataset is far less than the total population which is not sufficient to learn the entire speaker space. This directly results in poor generalization on the unseen speakers at inference leading to unsatisfactory performance.

The main challenge of ZSM-TTS is the speaker domain shift problem, which happens when the speaker outside of the training dataset must be inferred properly. It is commonly known that there is a strong bias towards the speakers from the training dataset. In order to overcome this challenge, we propose to train a TTS model with adversarial speaker-consistency learning (ASCL). The ASCL scheme generates an additional speech sample using a query speaker obtained from external untranscribed audio datasets. Such untranscribed audio datasets are readily available from various sources. The generated sample is then used for an adversarial training where the speaker-consistency discriminator is newly proposed. Training in this way exposes the model to a larger speaker pool than the limited training dataset, hence inducing better speaker generalization for ZSM-TTS.

The proposed method directly addresses to the aforementioned speaker domain shift problem by expanding the speaker pool for the ZSM-TTS. The ASCL scheme is built on the architecture of variational inference TTS (VITS) \cite{VITS} and its inverse transformation capability of normalizing flow module. We demonstrate the effectiveness of the ASCL by comparing with the baseline using subjective and objective scores. Our results show that the proposed method overcome the baseline in terms of speech quality and speaker similarity.

Our contributions are two-folded as follows:
\begin{enumerate}
    \item We propose adversarial speaker-consistency learning (ASCL), a novel way to train a TTS model to address the speaker domain shift problem.

    \item The proposed method leverages the external untranscribed audio datasets where such datasets are available from various sources to expose the TTS model to a larger speaker pool.
\end{enumerate}

\section{Background}
\subsection{VITS for multi-speaker TTS}
Our method extends the original VITS \cite{VITS} leveraging its invertible capability of normalizing flow module. VITS architecture is composed of a posterior encoder, a prior encoder, a decoder, a duration predictor, and a discriminator. The prior encoder is further divided into a text encoder and a normalizing flow which transforms the text-conditional prior distribution \(p_{\theta}(\textbf{\textit{z}}_{f}|\textbf{\textit{x}})\) into a more complex distribution \(p_{\theta}(\textbf{\textit{z}}_{v}|\textbf{\textit{x}})\). VITS combines the variational autoencoder formulation with an adversarial training scheme to successfully generate an audio in a phoneme-to-wav fashion without a separate vocoder training. VITS is trained to maximize the variational lower bound of a conditional log-likelihood
\begin{equation}
    \log p_{\theta}(\textbf{\textit{y}}|\textbf{\textit{x}}) \geq \mathbb{E}_{q_{\phi}(\textbf{\textit{z}}_{v}|\textbf{\textit{y}})}\left[\log p_{\theta}(\textbf{\textit{y}}|\textbf{\textit{z}}_{v})-\log \dfrac{q_{\phi}(\textbf{\textit{z}}_{v}|\textbf{\textit{y}})}{p_{\theta}(\textbf{\textit{z}}_{v}|\textbf{\textit{x}})}\right]_{\textstyle ,}
\end{equation}
where \(\textbf{\textit{x}}\), \(\textbf{\textit{y}}\), and \(\textbf{\textit{z}}_{v}\) denote the input phoneme sequence, the target waveform, and the latent acoustic embedding sequence respectively. The essential part of VITS training strategy is the distribution matching between the posterior distribution, \(q_\phi(\textbf{\textit{z}}_{v}|\textbf{\textit{y}})\), and the prior distribution, \(p_\theta(\textbf{\textit{z}}_{v}|\textbf{\textit{x}})\), via Kullback-Leibler divergence loss. The alignment between the text and audio sequence is learned via monotonic alignment search (MAS) algorithm originally proposed in \cite{Glow-TTS}.

In the multi-speaker setting, a speaker embedding vector \(g \in \mathbb{R}^{M}\) is conditioned to the normalizing flow \(f_\theta : \mathbb{R}^{D}\rightarrow{}\mathbb{R}^{D}\) where the text-conditional prior embedding sequence \(\textbf{\textit{z}}_{f} = f_\theta(\textbf{\textit{z}}_{v}, g)\) is trained to become speaker-independent via the forward transformation at training. At inference, VITS generates a speech using a speaker embedding vector \(g\) and the text-conditional prior embedding sequence \(\textbf{\textit{z}}_{f}\), exploiting the inverse transformation of the normalizing flow as follows:
\begin{align}
    \tilde{\textbf{\textit{y}}} = Dec_{\theta}(f^{-1}_\theta(\textbf{\textit{z}}_{f}, g))_{\textstyle ,}
\end{align}
where \(Dec_\theta\) denote the decoder.

\section{Proposed method}
Our method builds on top of the original VITS architecture and training methodology. The proposed method resolves the speaker domain shift problem with a two-stage approach. Firstly, we generate a speech sample of a query speaker at each training iteration utilizing the external untranscribed audio datasets such that the TTS model is exposed to an extensively large speaker set. Secondly, a speaker-consistency discriminator is introduced to determine if the generated speech sample follows the speaker identity of the given speaker embedding vector in an adversarial manner.

\subsection{Speech generation from a query speaker}
At each training iteration, we randomly draw a text and a waveform pair from a TTS dataset \([\textbf{\textit{x}}^{t}, \textbf{\textit{y}}^{t}_{s}]\) as a support set and another waveform from an external untranscribed audio dataset \([\textbf{\textit{y}}^{u}_{q}]\) as a query to form an input tuple \([\textbf{\textit{x}}^{t}, \textbf{\textit{y}}^{t}_{s}, \textbf{\textit{y}}^{u}_{q}]\) where the speaker sets of the TTS dataset and the untranscribed dataset are disjoint. Our goal is to generate an output pair \([\tilde{\textbf{\textit{y}}}^{t}_{s}, \tilde{\textbf{\textit{y}}}^{t}_{q}]\) where it is evaluated by two different objectives. At training, a posterior encoder first receives \(\textbf{\textit{y}}^{t}_{s}\) as an input and outputs a latent acoustic embedding sequence \({\textbf{\textit{z}}_{v}}^{t}_{s}\). \({\textbf{\textit{z}}_{v}}^{t}_{s}\) then goes through the decoder \(Dec_\theta\) to generate \(\tilde{\textbf{\textit{y}}}^{t}_{s}\) in an autoencoding fashion. 
On the other hand, we can also generate \(\tilde{\textbf{\textit{y}}}^{t}_{q}\) which contains the content of \(\textbf{\textit{x}}^{t}\) and the speaker identity of \(\textbf{\textit{y}}^{u}_{q}\) by exploiting the forward and inverse transformation of the flow module with the speaker embedding vectors \(g_s, g_q \in \mathbb{R}^{M}\) extracted from \(\textbf{\textit{y}}^{t}_{s}\) and \(\textbf{\textit{y}}^{u}_{q}\) respectively as follows:
\begin{align}
    \tilde{\textbf{\textit{y}}}^{t}_{q} &= Dec_\theta(f^{-1}_\theta(f_\theta({\textbf{\textit{z}}_{v}}^{t}_{s}, g_s), g_q))_{\textstyle .}
\end{align}

The motivation behind using the external untranscribed audio data is to extensively expand the speaker set of the training data where only a relatively small number of speakers are available within TTS datasets due to the requirement of the high-quality pairing of \([text, audio]\). By utilizing the untranscribed datasets which are easily accessible from various sources including but not limited to speaker verification and automatic speech recognition datasets, the number of speakers that we gain becomes enormously larger than that of TTS datasets. This enforces the TTS model to be exposed to a more diverse speaker pool at training and enhances the overall speaker generalization.

\subsection{Adversarial learning}
Since the ground truth \(\textbf{\textit{y}}^{t}_{q}\) does not exist, the reconstruction loss of VITS cannot be imposed on \(\tilde{\textbf{\textit{y}}}^{t}_{q}\). In order to circumvent this problem, an adversarial learning technique is employed to evaluate on the generated output pair \([\tilde{\textbf{\textit{y}}}^{t}_{s}, \tilde{\textbf{\textit{y}}}^{t}_{q}]\) and their corresponding speaker embedding vectors \(g_s, g_q\). We introduce a speaker-consistency discriminator, \(D_{\omega}\), which takes a waveform and a speaker embedding vector \([\textbf{\textit{y}}, g]\) as an input pair. \(D_{\omega}\) tries to determine whether the input waveform is consistent with the given speaker embedding in terms of speaker identity. Utilizing the generated outputs and ground truth samples, \(D_\omega\) distinguishes the real pairs \([\textbf{\textit{y}}^{u}_{q}, g_q], [\textbf{\textit{y}}^{t}_{s}, g_s]\) from the generated pairs \([\tilde{\textbf{\textit{y}}}^{t}_{q}, g_q], [\tilde{\textbf{\textit{y}}}^{t}_{s}, g_s]\). The adversarial objective \(\mathcal{L_{ASCL}}\) for the speaker-consistency discriminator is formulated as follows:

\begin{equation}
\resizebox{0.91\columnwidth}{!}{
$
\begin{aligned}
    \mathcal{L_{ASCL}}(D_{\omega})&=\mathop{\mathbb{E}}_{\substack{g_s, g_q,\textbf{\textit{z}}_{f}^{t}}}[\alpha(D_{\omega}(\textbf{\textit{y}}^{u}_{q}, g_q)-1)^2 + (D_{\omega}(\textbf{\textit{y}}^{t}_{s}, g_s)-1)^2\\
    &\qquad+\alpha(D_{\omega}(\tilde{\textbf{\textit{y}}}^{t}_{q},g_q))^2+(D_{\omega}(\tilde{\textbf{\textit{y}}}^{t}_{s},g_s))^2]_{\textstyle ,}
\end{aligned}
$
}
\end{equation}
\begin{equation}
\resizebox{0.91\columnwidth}{!}{
$
\begin{aligned}
    \mathcal{L_{ASCL}}(G_{\theta}) = \mathop{\mathbb{E}}_{\substack{g_s, g_q,\textbf{\textit{z}}_{f}^{t}}}[\alpha(D_{\omega}(\tilde{\textbf{\textit{y}}}^{t}_{q}, g_q)-1)^2 + (D_{\omega}(\tilde{\textbf{\textit{y}}}^{t}_{s}, g_s)-1)^2]_{\textstyle ,}
\end{aligned}
$
}
\end{equation}
where we set \(\alpha=0.3\).

\(G_\theta\) is the generator part of VITS architecture in this case, where \(G_\theta(\textbf{\textit{z}}_{f}, g) = Dec_\theta(f^{-1}_\theta(\textbf{\textit{z}}_{f}, g))=\tilde{\textbf{\textit{y}}}\). The \(\mathcal{L_{ASCL}}\) utilizes LS-GAN \cite{LS-GAN} instead of the original GAN \cite{GAN} loss for stable training.

\begin{figure}
    \centering
    \includegraphics[width=\linewidth]{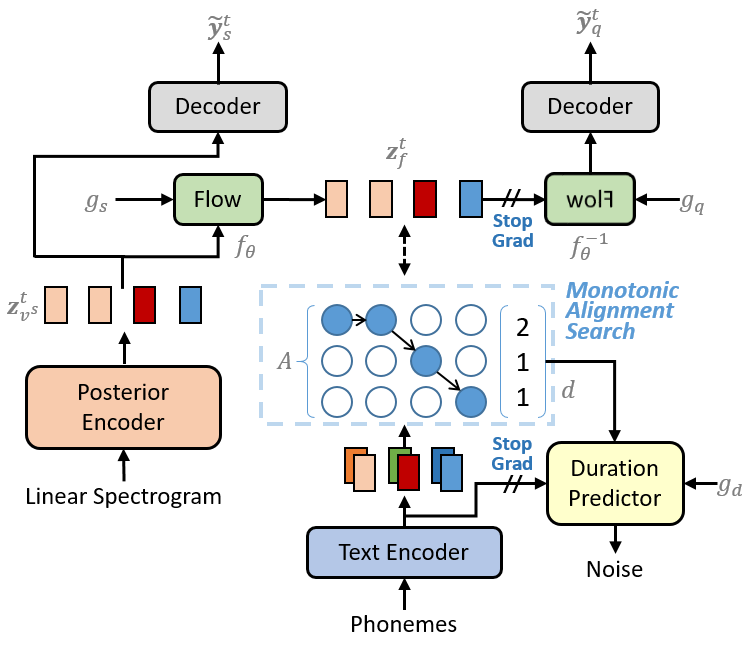}
    \caption{The architecture of ASCL-VITS scheme and the training pipeline.}
    \label{fig:training}
\end{figure}

To prevent the ASCL from affecting the training of unrelated modules such as posterior encoder, text encoder, and duration predictor, we use a stop gradient operator at \({\textbf{\textit{z}}_{f}^{t}}\) to restrain the back-propagation of the gradient flow. The training pipeline of ASCL built on VITS (ASCL-VITS) is shown in Fig. \ref{fig:training}.

\subsection{Speaker-consistency discriminator}
\begin{figure}
    \centering
    \includegraphics[width=0.75\linewidth]{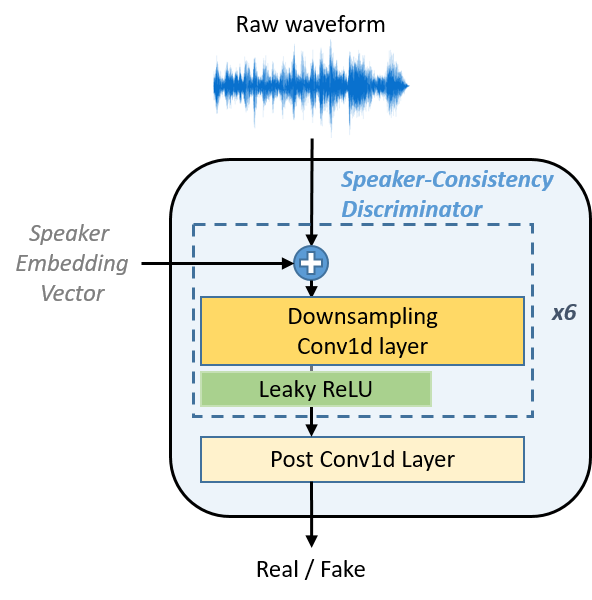}
    \caption{The block diagram of the speaker-consistency discriminator.}
    \label{fig:discriminator}
\end{figure}

We modify the original multi-scale discriminator (MSD) architecture, proposed in MelGAN \cite{MelGAN}, with a speaker embedding vector \(g\) as a conditional input. Unlike the original MSD where the discriminator is a mixture of three sub-discriminators which takes an audio segment on three different scales, our speaker-consistency discriminator \(D_\omega\) operates on the raw audio scale without any sub-discriminators. \(D_\omega\) is composed of a stack of six 1-D strided and grouped convolutional layers each followed by leaky ReLU activations \cite{leakyReLU}. Each convolutional layer performs a downsampling operation with the kernel size of 4, which captures the features from the smoothed waveform at different scales. A speaker embedding vector is added to the input of each convolutional layer to evaluate the speaker consistency at each downsampling stage. A post 1-D convolutional layer with the kernel size of 3 is added at the end of the stack to produce the output. The detailed block diagram of \(D_\omega\) is shown in Fig. \ref{fig:discriminator}.

\section{Experiments and results}
\subsection{Implementation details}
The proposed method adds on to the original VITS using the official implementation code\footnote{\url{https://github.com/jaywalnut310/vits}} with a few modifications.

\subsubsection{Speaker encoder module}
In order to extract the speaker embedding vectors of a large set of speakers, a pre-trained speaker encoder from speaker verification task is employed. We used the Fast ResNet-34 model with a contrastive equilibrium learning (CEL) \cite{CEL} trained on VoxCeleb2 \cite{VoxCeleb2}, and its official implementation can be found here\footnote{\url{github.com/msh9184/contrastive-equilibrium-learning}}. We trained the speaker encoder with the input of 80-dimensional log mel-spectrograms to extract a 512-dimensional speaker embedding vector. A linear layer which reduces the embedding dimension to 256 followed by a ReLU activation and a linear projection is added to the pre-trained speaker encoder to obtain the final speaker embedding vector. The weights of the pre-trained speaker encoder are frozen throughout the training and inference of ASCL-VITS to consistently draw the speaker embedding vectors from a single learned speaker embedding space.

\subsubsection{VITS speaker conditioning}
While conditioning the speaker embedding vector to all submodules is known to enhance the overall TTS performance, we restrict our VITS model to condition the speaker embedding to the normalizing flow module and the duration predictor to evaluate the proposed method effectively. Moreover, since the duration is more related to the speech rate of the reference given at inference as well as the identity of the speaker, we train a separate reference encoder to extract an utterance-level reference embedding vector \(g_d\) to condition the duration predictor. The reference encoder follows the same architecture as that of the global style token \cite{GST} with the output dimension of 256.

\subsection{Experimental settings and datasets}
\subsubsection{Datasets}
For ZSM-TTS evaluation, we used the VCTK \cite{VCTK} dataset, which consists of 108 speakers and contains approximately 44 hours of speech recording. We selected 11 speakers as an in-domain test set following \cite{SC-GlowTTS} and \cite{YourTTS}, and used the remaining 97 speakers as a paired training dataset. We combined the audio data of LibriTTS-clean-100, LibriTTS-clean-360, and LibriTTS-other-500 from LibriTTS dataset \cite{LibriTTS} to build an untranscribed dataset for training. The combined dataset contains 2,311 speakers with approximately 554 hours of speech recording. LibriTTS-test-clean dataset, which is composed of 39 speakers with 9.56 hours of speech recording, is used for the out-of-domain evaluation. We downsampled the audio to 24kHz for training and inference. We extract linear spectrograms with a 1024 Fast Fourier Transform (FFT) size, 256 hop size, and 1024 window size. Then, 80-dimensional mel-scale filterbank is applied to convert the linear spectrograms to mel-spectrograms.

\subsubsection{Experimental setup}
We first built a VITS model with the proposed ASCL scheme, referred as \textbf{VITS-ASCL}. To demonstrate the effectiveness of ASCL, we trained a vanilla VITS model with the aforementioned pre-trained speaker encoder, referred as \textbf{Multi-speaker VITS}. This baseline has the same architecture as \textbf{ASCL-VITS} without ASCL.

All models were trained using the training set of VCTK which is composed of 97 speakers. To evaluate the baseline and proposed models, we conducted the experiments in two different settings. First, we generated the speech samples from the remaining 11 speakers of VCTK dataset to evaluate on in-domain unseen speakers. For the second experimental setup, we drew 20 speakers randomly from LibriTTS test-clean dataset and generated speech samples to evaluate on out-of-domain unseen speakers. We randomly selected 25 samples from each model depending on the measure and experimental setup.

\subsubsection{Evaluation method}
To evaluate the effectiveness of ASCL, we first conducted a subjective test to measure the quality of the generated speech samples from each model with mean opinion score (MOS). We also compared the generated speech samples from each model in terms of the speaker similarity via measuring similarity mean opinion score (SMOS). Both measures are 5-point scale ranging from 1 to 5. The participants were asked to evaluate on the quality of samples in terms of intelligibility and naturalness for MOS test, whereas participants were asked to rate the speaker similarity between the given speech sample and the ground truth sample for SMOS test. For the evaluation, 14 participants were asked to rate MOS and SMOS tests on VCTK and LibriTTS unseen speakers. The resulting scores are presented with 95\% confidence intervals in the Table \ref{table:1}.

Along with the above subjective tests, we also conducted an objective test which measures the cosine distance similarity between the two speaker embedding vectors extracted from the generated sample and the ground truth sample, respectively. This score, also known as speaker embedding cosine similarity (SECS), ranges from -1 to 1, and as the score is closer to 1, the two given speaker embedding vectors are closer in terms of speaker similarity. We used a publicly available toolkit, SpeechBrain\footnote{\url{https://speechbrain.github.io/}} \cite{SpeechBrain} which includes a speaker encoder trained on a speaker verification task.

\setlength{\tabcolsep}{5pt}
\begin{table} [h]
    \renewcommand{\arraystretch}{1.2}
    \centering
    \captionsetup{justification=centering}
    \caption{MOS and SMOS on VCTK and LibriTTS unseen speakers}
    \begin{tabular}{c c c || c c}
        \toprule
        \multirow{2}{*}{\textbf{Model}}&\multicolumn{2}{c}{\textbf{VCTK}}&\multicolumn{2}{c}{\textbf{LibriTTS}}\\
        \cline{2-5} 
        & \textbf{MOS}($\uparrow$) & \textbf{SMOS}($\uparrow$) & \textbf{MOS}($\uparrow$) & \textbf{SMOS}($\uparrow$)\\
        
        \specialrule{0.1pt}{1pt}{1pt}
        Ground Truth & 4.76$\pm$0.03 & 4.68$\pm$0.03 & 4.62$\pm$0.03 & 4.47$\pm$0.04\\
        \specialrule{0.1pt}{1pt}{1pt}
        Multi-speaker VITS & 4.20$\pm$0.04 & 3.81$\pm$0.05 & 4.01$\pm$0.04 & 3.11$\pm$0.06 \\
        \textbf{ASCL-VITS} & \textbf{4.50$\pm$0.03} & \textbf{4.11$\pm$0.04} & \textbf{4.37$\pm$0.04} & \textbf{3.52$\pm$0.05}\\
        \bottomrule
    \end{tabular}
    \label{table:1}    
\end{table}

\setlength{\tabcolsep}{10pt}
\begin{table} [h]
    \renewcommand{\arraystretch}{1.2}
    \centering
    \captionsetup{justification=centering}
    \caption{SECS comparison between the baseline and proposed model on VCTK and LibriTTS unseen speakers}
    \begin{tabular}{c c || c}
        \toprule
        \multirow{2}{*}{\textbf{Model}}&\multicolumn{2}{c}{\textbf{SECS}($\uparrow$)}\\
        \cline{2-3} 
        & \textbf{VCTK} & \textbf{LibriTTS}\\
        
        \specialrule{0.1pt}{1pt}{1pt}
        Ground Truth & 0.78 & 0.70\\
        \specialrule{0.1pt}{1pt}{1pt}
        Multi-speaker VITS & 0.35 & 0.19 \\
        \textbf{ASCL-VITS} & \textbf{0.37} & \textbf{0.23}\\
        \bottomrule
    \end{tabular}
    \label{table:2}    
\end{table}

\subsection{Results}
As presented in Tables \ref{table:1} and \ref{table:2}, our proposed method, \textbf{ASCL-VITS}, demonstrates the effectiveness with higher scores in MOS, SMOS, and SECS on both VCTK and LibriTTS unseen speaker. One notable observation is that the \textbf{Multi-speaker VITS} encountered multiple occasions where the speaker identity is switched within a generated sample in LibriTTS unseen speaker experiment. However, we did not observe such phenomenon from the samples generated from \textbf{ASCL-VITS}. This shows that our method is more robust in terms of speaker generalization in out-of-domain cases compared to the baseline.
The overall results show that ASCL is an effective and suitable method to enhance the speaker generalization of a TTS model for the ZSM-TTS task.

\section{Conclusions}
In this paper, adversarial speaker-consistency learning (ASCL) is proposed to mitigate the speaker domain shift problem of ZSM-TTS. We first generate a speech of a query speaker using an extensive speaker set from untranscribed datasets and learns to generate the samples of an unseen speaker via an adversarial learning with the speaker-consistency discriminator. The ASCL-applied VITS model outperforms the baseline in ZSM-TTS with both in-domain and out-of-domain experimental settings in terms of the quality and the speaker similarity. The experimental results demonstrate that ASCL is effective in synthesizing high-quality speech samples given a reference audio from an unseen speaker. For future works, we further develop ASCL by extending to different speech factors, such as emotion, prosody, and dialect by utilizing generative models for a more diverse zero-shot multi-factor text-to-speech scenarios.

\section*{Acknowledgment}
This work was supported by Institute of Information \& communications Technology Planning and Evaluation (IITP) grant funded by the Korea government(MSIT) (No.2021-0-00456, Development of Ultra-high Speech Quality Technology for Remote Multi-speaker Conference System)


\begin{thebibliography}{14}

\bibitem{Tacotron2}J. Shen {\em et al.}, "Natural TTS synthesis by conditioning wavenet on mel spectrogram predictions," in {\em Proc. IEEE Int. Conf. Acoust., Speech, Signal Process.}, Calgary, AB, Canada, 15–20 April 2018, pp. 4779–4783.

\bibitem{Wavenet} A. Oord et al., "Wavenet: A generative model for raw audio," 2016, {\em arXiv preprint arXiv:1609.03499}.

\bibitem{Transfer-Tacotron}R. Skerry-Ryan {\em et al.}, “Towards end-to-end prosody transfer for expressive speech synthesis with Tacotron,” in {\em Proc. Int. Conf. Mach. Learn.}, 2018, pp. 4693–4702.

\bibitem{zero-shot-tts} E. Cooper et al., "Zero-shot multi-speaker text-to-speech with state-of-the-art neural speaker embeddings," in {\em Proc. IEEE Int. Conf. Acoust., Speech, Signal Process.}, Barcelona, Spain, 4–8 May 2020, pp. 6184-6188.

\bibitem{SC-GlowTTS} E. Casanova et al., "SC-GlowTTS: an efficient zero-shot multi-speaker text-to-speech model," 2021, {\em arXiv:2104.05557}.

\bibitem{YourTTS} E. Casanova, J. Weber, C. Chulby, A. Junior, E. Golge, and M. Ponti, "YourTTS: Towards zero-shot multi-speaker TTS and zero-shot voice conversion for everyone," 2022, \emph{arXiv:2112.02418}.

\bibitem{GST}Y. Wang {\em et al.}, "Style tokens: Unsupervised style modeling, control and transfer in end-to-end speech synthesis," in {\em Proc. Int. Conf. Mach. Learn.}, Stockholm, Sweden, 10–15 July 2018, Volume 80, pp. 5180–5189.

\bibitem{Meta-StyleSpeech} D. Min, D. Lee, E. Yang, and S. Hwang, "Meta-StyleSpeech: Multi-speaker adaptive text-to-speech generation," in {\em Proc. Int. Conf. Mach. Learn.,} 2021, pp. 7748-7759.

\bibitem{VITS} J. Kim, J. Kong, and J. Son, "Conditional variational autoencoder with adversarial learning for end-to-end text-to-speech," in {\em Proc. Int. Conf. Mach. Learn.,} 2021, pp. 5530-5540.

\bibitem{Glow-TTS} J. Kim, S. Kim, J. King, and S. Yoon, "Glow-TTS: A generative flow for text-to-speech via monotonic alignment search," in {\em proc. Neural Inf. Process. Syst.}, vol. 33, 2020, pp. 8067-8077.

\bibitem{MelGAN} K. Kumar et al., "MelGAN: Generative adversarial networks for conditional waveform synthesis," in {\em proc. Neural Inf. Process. Syst.}, 2019.

\bibitem{leakyReLU} B. Xu, N. Wang, T. Chen, and M. Li, "Empirical evaluation of rectified activations in convolution network," 2015, \emph{arXiv:1505.00853}.

\bibitem{CEL} S. Mun, W. Kang, M. Han, and N. Kim, "Unsupervised representation learning for speaker recognition via contrastive equilibrium learning," 2020, \emph{arXiv:2010.11433}.

\bibitem{VoxCeleb2} J. Chung, A. Nagrani, and A. Zisserman, "VoxCeleb2: Deep speaker recognition,” in {\em Proc. Interspeech}, Hyderabad, India,2-6 Sep 2018, pp. 1086–1090.

\bibitem{VCTK} J. Yamagishi, C. Veaux, and K. MacDonald, "CSTR VCTK corpus: English multi-speaker corpus for CSTR voice cloning toolkit (version 0.92), 2019.
\balance

\bibitem{LibriTTS} H. Zen {\em et al.}, "LibriTTS: A corpus derived from LibriSpeech for text-to-speech," in {\em Proc. Interspeech}, Graz, Austria, 16–19 Sept 2019, pp. 1526–1530.

\bibitem{Tacotron}Y. Wang {\em et al.}, "Tacotron: Towards end-to-end speech synthesis," in {\em Proc. Interspeech}, Toronto, ON, Canada, 24–28 June 2017, pp. 4006–4010.

\bibitem{H/ASP} H. Heo, B. Lee, J. Huh, and J. Chung, “Clova baseline system for the voxceleb speaker recognition challenge 2020,” 2020, {\em arXiv preprint arXiv:2009.14153}.

\bibitem{SpeechBrain} M. Ravanelli et al., “Speechbrain: A general-purpose speech toolkit,” 2021, {\em arXiv preprint arXiv:2106.04624}.

\bibitem{adam} D. Kingma and J. Ba, "Adam: A method for stochastic optimization," in {\em Proc. Int. Conf. Learn. Representation}, 2015.

\bibitem{VAE} D. Kingma and M. Welling, "Auto-encoding variational bayes," in {\em Proc. Int. Conf. Learn. Representation}, 2014.

\bibitem{NICE} L. Dihn, D. Krueger, and Y. Bengio, "NICE: Non-linear independent components estimation," 2015, \emph{arXiv:1410.8516}.

\bibitem{RealNVP} L. Dihn, J. Solh-Dickstein, and S. Bengio, "Density estimation using real NVP," in {\em Proc. Int. Conf. Learn. Representation}, 2015.

\bibitem{EATS} J. Donahue, S. Dieleman, M. Binkowski, E. Elsen, and K. Simonyan, "End-to-end adversarial text-to-speech," in {\em arXiv preprint arXiv:2006.03575}, 2020.

\bibitem{Hifigan} J. Kong, J. Kim, and J. Bae, “Hifi-gan: Generative adversarial networks for efficient and high fidelity speech synthesis,” in {\em Proc. Adv. Neural Inf. Process. Syst.}, vol. 33, 2020, pp. 17022-17033.

\bibitem{14}R. Skerry-Ryan {\em et al.}, “Towards end-to-end prosody transfer for expressive speech synthesis with Tacotron,” in {\em Proc. Int. Conf. Mach. Learn.}, 2018, pp. 4693–4702.

\bibitem{Multi-SpectroGAN} S. Lee, H. Yoon, H. N, J. Kim, and S. Lee, "Multi-SpectroGAN: High-diversity and high-fidelity spectrogram generation with adversarial style combination for speech synthesis," in {\em Proc. AAAI Conf. Artif. Intell.}, 2021, pp. 13198-13206.

\bibitem{Normalization-Zero-shot} N. Kumar, S. Goel, A. Narang, and B. Lall, "Normalization driven zero-shot multi-speaker speech synthesis," in {\em Proc. Interspeech}, Brno, Czechia, 30 August – 3 September, 2021, pp. 1354–1358.

\bibitem{LS-GAN} X. Mao et al., "Least squares generative adversarial networks," in {\em Proc. Int. Conf. Comp. Vision,}, Venice, Italy, 2017, pp.2813-2821.

\bibitem{GAN} I. Goodfellow et al., "Generative adversarial nets," in {\em Proc. Adv. Neural Inf. Process. Syst.}, vol. 27, 2014.

\bibitem{Adversarial_zsl} B. Tong et al., "Adversarial zero-shot learning with semantic augmentation," in {\em Proc. AAAI Conf. Artif. Intell.}, 2018, pp. 2476-2483.

\bibitem{zero-vae-gan} R. Gao et al., "Zero-VAE-GAN: Generating unseen features for
generalized and transductive zero-shot learning," {\em IEEE Trans. on Image Proc.}, vol 29, 2020, pp. 3665-3680.

\bibitem{Flow-GAN} A. Grover, M. Dhar, and S. Ermon, "Flow-GAN: Combining maximum likelihood and adversarial learning in generative models," in {\em Proc. AAAI Conf. Artif. Intell.}, 2018, pp. 3069-3076.

\bibitem{Alignflow} A. Grover, C. Chute, R. Shu, Z. Cao, and S. Ermon, "Aligflow: Cycle consistent learning from multiple domains via normalizing flows," in {\em Proc. AAAI Conf. Artif. Intell.}, 2018, pp. 4028-4035.

\end{thebibliography}
\end{document}